\documentclass[english]{ieej-e}
\usepackage[dvips]{graphicx}
\usepackage[varg]{txfonts}

\FIELD{}
\YEAR{}
\NO{}
\title{Sensorless Measurement of Solenoid Stroke and Temperature using Convolution Neural Network with Two Points of PWM Driving Current}
\authorlist{%
 \authorentry[akita@is.t.kanazawa-u.ac.jp]{Junichi Akita}{n}{KU}
}
\affiliate[KU]{Kanazawa University \\ Kakuma, Kanazawa, 920-1192 Japan\\
}

\newcommand\degC{${}^\circ$C}
\newcommand\Ton{$T_{\rm ON}$}

\begin{document}

\begin{abstract}
In this paper, we describe the algorithm to measure the stroke and the temperature of solenoid using PWM driving current at two points based on the electric characteristics of the solenoid with CNN, without mechanical attachments.
We describe the evaluation experimental results of the stroke and the temperature prediction.
We also describe the preliminary experimental results of controlling the solenoid stroke at intermediate position.
\end{abstract}


\maketitle

\section{Introduction}

Solenoid is the actuator composed of the coil and the plunger.
Solenoid is used to push or pull the object, and the stroke is controlled in two states; ON and OFF, in usual cases.
From the viewpoint of electrical characteristics of the solenoid, its inductance changes according to the plunger position, and we can measure the plunger position using the electrical characteristics of the solenoid.
We've already reported the method of measuring the solenoid stroke (plunger's position) by using electric characteristics, without mechanical attachments\cite{sol_arxiv}.
However, it has restrictions on the target solenoid's electric characteristics, and the solenoid where it can be applied.

In this paper, we propose to method to measure the stroke and the temperature of the solenoid using convolution neural network (CNN) for two measured values of driving current, as well as its evaluation on the accuracy.
We also describe the preliminary experimental results of the plunger's position control without mechanical attachments.

\section{Related Works}

\begin{figure}[t]
{\hfill
\includegraphics[width=0.6\columnwidth]{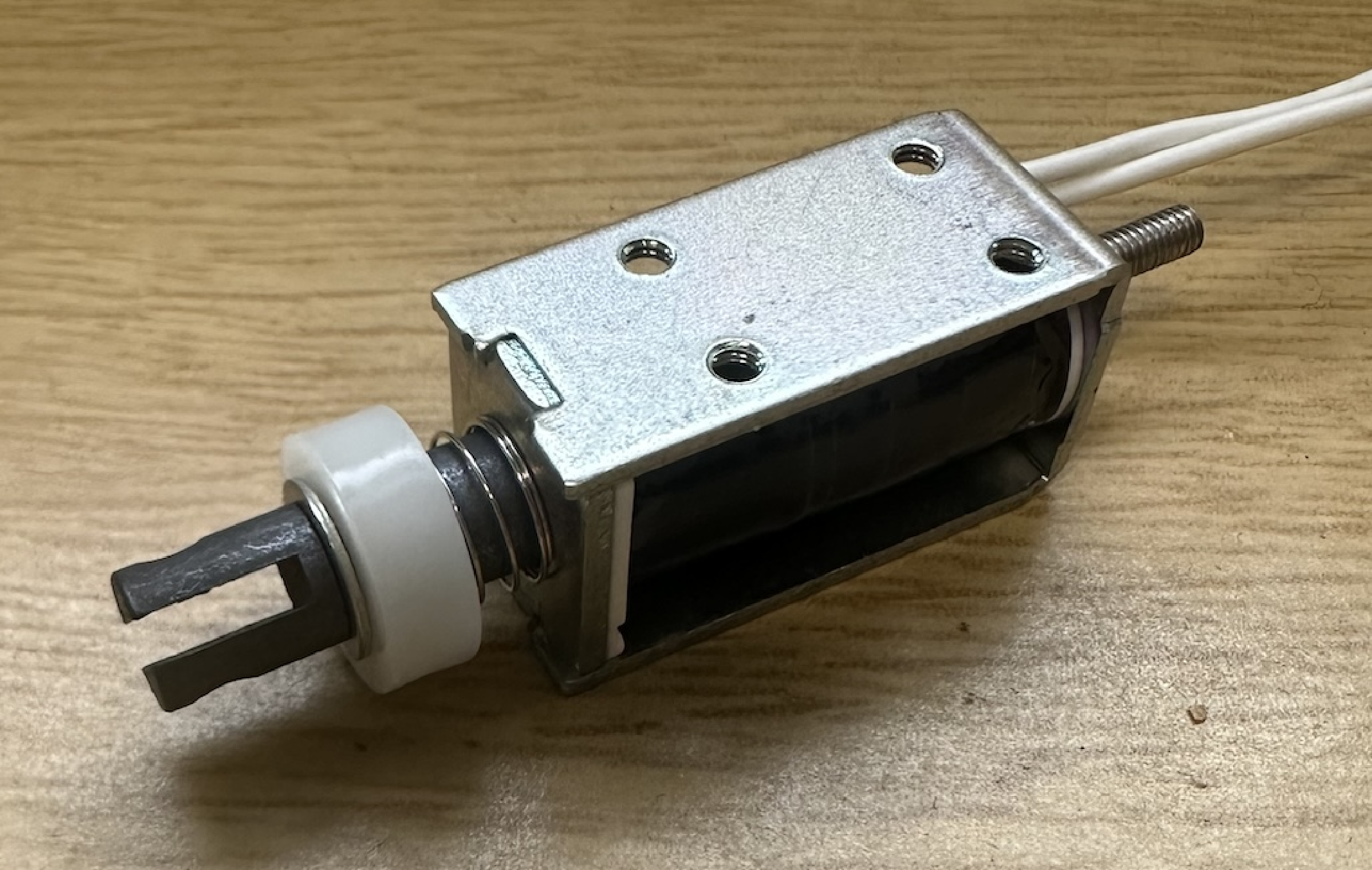}
\hfill}
\caption{Solenoid (Takaha's CBS07300580)}
\label{fig:solenoid}
\end{figure}

Solenoid is the actuator composed of the coil and the plunger, as shown in Fig.\ref{fig:solenoid}.

There are two different types of the solenoid.
One is the ON-OFF solenoid, or the conventional solenoid, which has two states of ON and OFF.
The other is the proportional solenoid, whose plunger's position is proportional to the current of the coil, which is often used for liquid or air flow control\cite{flowcontrol}.
Since its plunger's position is determined by the equilibrium of the coil's force and the plunger's mass gravity, the external force may affect the error from the ideal position of the plunger.
There are researches on the detailed modeling of the proportional solenoid\cite{solenoid1,solenoid2}.

There is the motivation to use the ON-OFF solenoid as the proportional solenoid for reducing cost and system complexity.
It is realized by controlling the solenoid's plunger at intermediate positions.
Some of them utilize the external sensor to measure the plunger's position\cite{control1,control2,control2}, while some describe the sensorless measurement methods.

There are several approaches on the sensorless measurement of the solenoid's plunger position.
They basically measure the change of the inductance of the solenoid on the plunger's position.
M.F.Rahman utilizes the incremental inductance's change on the plunger's position, with measuring the rising current of the solenoid's coil with PWM drive\cite{measure1}.
D.Pawelczak uses the real part of the solenoid's impedance\cite{measure3}.
J.C.Renn utilizes the slope of the current on PWM drive\cite{measure4}.
I.Duelk utilizes the average of the current of the solenoid's coil with sinusoid-modulated PWM drive\cite{measure5}.
S.T.Wu utilizes the current phase of PWM drive\cite{measure8}.
F.Straussberger has proposed the method of determining the solenoid's system parameters based on the solenoid's model\cite{measure9}.
These researches utilize the time-domain current wave information to estimate the solenoid's plunger position, which requires the series of data acquisition.

B.Hanson measures the electromotive force for plunger's position\cite{measure7}, which is hard to measure with high accuracy.

S.Nagai has proposed the overlaying AC sinusoid wave on PWM driving pulse to measure the inductance of the solenoid\cite{measure2}.
J.Maridor utilizes the resonant frequency change on the plunger's position\cite{measure6}.
These researches utilize the frequency-domain information, which has difficulties on real-time measurement.

\section{Electrical Characteristics of Solenoid with PWM Drive}

\begin{figure}[t]
{\hfill
\includegraphics[width=0.3\columnwidth]{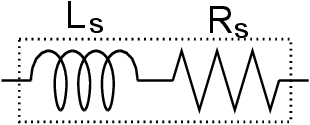}
\hfill}
\caption{Equivalent circuit of solenoid}
\label{fig:solenoid-circuit}
\end{figure}

From the viewpoint of electrical characteristics of the solenoid, it is a simple inductor, whose equivalent electrical circuit is described as shown in Fig.\ref{fig:solenoid-circuit}, where $L_s$ and $R_s$ are the inductance and the wire resistance, respectively.

\begin{figure}[t]
{\hfill
\includegraphics[width=0.9\columnwidth]{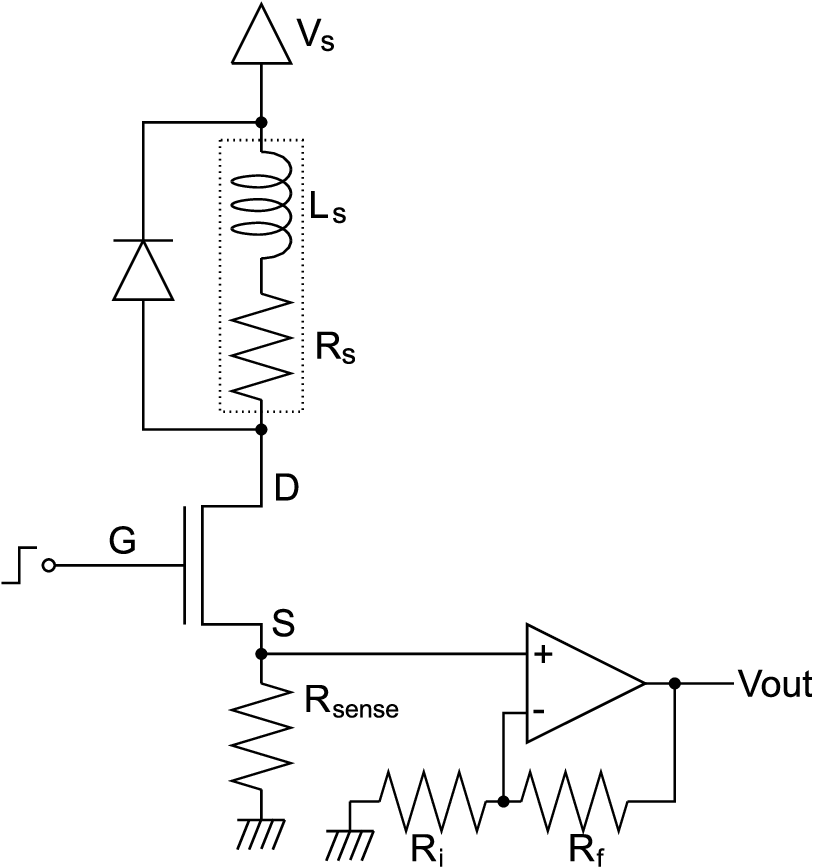}
\hfill}
\caption{Driving circuit of solenoid}
\label{fig:drive-circuit}
\end{figure}

The typical driving circuit of the solenoid is shown in Fig.\ref{fig:drive-circuit}.
In PWM drive, usually used for driving the solenoid, there are two states as follows.
\begin{itemize}
\item ON: giving a high voltage to G (gate) to turn the nMOSFET on, where the current flows to make the plunger pulled.
\item OFF: giving a low voltage to G to turn the nMOSFET off, where the current stops to make the plunger released.
\end{itemize}
The transitional phenomenon occurs between switching OFF-to-ON and ON-to-OFF, and can be theoretically analyzed\cite{sol_arxiv} as the current waveform depending on the inductance, $L$, and the PWM duty ratio, $r$, the ratio of ON time over the pulse cycle time.
It is notable that the current waveform depends not only the inductance, but also PWM duty ratio.

The plunger in the solenoid is a magnetic material, and it affects the inductance of solenoid, and the inductance, $L$ of solenoid is basically proportional to the position of the plunger\cite{sol_arxiv}.
We can determine the inductance, $L$, from the current waveform for given PWM duty ratio, $r$, and thus, we can determine the solenoid stroke.

We can measure the driving current during ON state as the voltage of $R_{\rm sense}$ in Fig.\ref{fig:drive-circuit} under the assumption of the voltage of $R_{\rm sense}$ is negligible.
The non-inverting amplifier in Fig.\ref{fig:drive-circuit} amplifiers the voltage of $R_{\rm sense}$ by manification of $(R_f + R_i)/R_i$. to obtain the output of $V_{\rm out}$.

\section{Stroke Measurement using Convolution Neural Network}

The change of the current waveform for different $L$ and $r$ is complicated.
From the viewpoint of the practical application, the full-wave measurement of the current waveform using A/D converter and microcontroller is difficult in the small microcontrollers.

\begin{figure}[t]
{\hfill
\includegraphics[width=0.7\columnwidth]{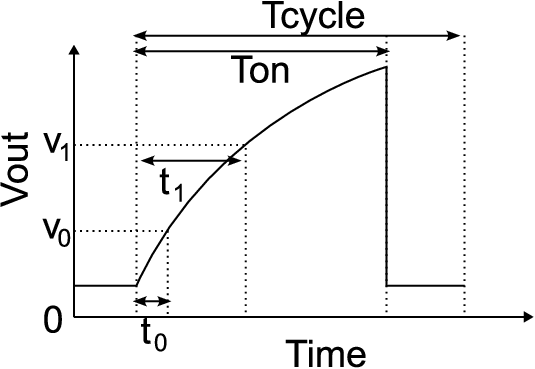}
\hfill}
\caption{Two sampled values in the current waveform of the solenoid drive}
\label{fig:waveform-sample}
\end{figure}

Here we try to predict the solenoid stroke, $x$ from two sampled values of the current waveform, $v_0$ and $v_1$, as shown in Fig.\ref{fig:waveform-sample}.
\Ton\ and $T_{\rm cycle}$ are ON time and teh cycle time of driving PWM signal, respectively.
$t_0$ and $t_1$ are fixed time from the starting edge of PWM ON signal to sampling point of the voltage values, $v_0$ and $v_1$ of the current sense amplifier in Fig.\ref{fig:drive-circuit}, respectively.

Since the PWM duty ratio, $r$, is determined by the driving controller, we can use $r$ as the known parameter to predict the solenoid stroke.
The inductance of the solenoid changes for its temperature, $T$, and the prediction of $T$ is also needed.

\begin{figure}[t]
\begin{center}
\includegraphics[width=0.7\columnwidth]{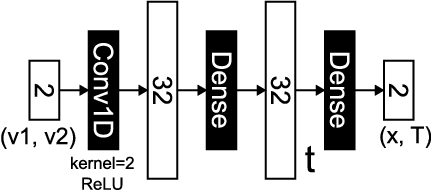}
\end{center}
\caption{The architecture of CNN used in this experiment}
\label{fig:cnn-structure}
\end{figure}

So, we define the problem as to predict $(x, T)$ from $(v_0, v_1, r)$.
We use the convolution neural network (CNN) to solve this problem.
Figure \ref{fig:cnn-structure} shows the architecture of CNN to solve this problem.

We defined the PWM cycle time as 10[ms] (PWM frequency of 100Hz), and the sampling timing $t_0$ and $t_1$ as 100[$\mu$s] and 700[$\mu$s], respectively.
Here we assume PWM duty cycle $r$ between 10[\%] and 90[\%], thus PWM ON state time between 1000[$\mu$s] and 9000[$\mu$s] to satisfy the conditions of $v_1$ sampling during ON state, and both ON-OFF and OFF-ON state in PWM cycle occur.

\begin{table}[t]
\caption{Used solenoids and the drive circuit parameters}
\label{tab:solenoids}
\begin{center}
\begin{tabular}{|c|c|c|c|}\hline
Type                      & Stroke[mm] & $V_s$[V] & $R_f[{\rm k}\Omega]$ \\ \hline
CBS07300380\cite{CBS0730} & 5          & 6        & 38   \\
CB10370140\cite{CB1037}   & 10         & 12       & 38   \\
SSBH-0830-01\cite{SSBH}   & 5          & 5        & 19.4 \\ \hline
\end{tabular}
\end{center}
\end{table}

Table \ref{tab:solenoids} shows three solenoids used in this experiment.
The supply voltage, $V_s$ is determined from the solenoid's specifications, and the feedback resistor of the amplifier, $R_f$, is determined from the actual ON current to keep the working range of the operational amplifier with $R_i=1[{\rm k}\Omega]$ and $R_{\rm sense}=0.2[\Omega]$.

\begin{figure}[t]
{\hfill
\includegraphics[width=0.9\columnwidth]{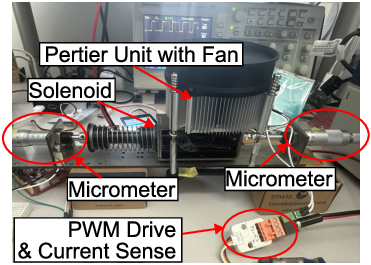}
\hfill}
\caption{Experimental setup}
\label{fig:setup}
\end{figure}

We used three solenoids in Tab.\ref{tab:solenoids} for experiments.
Figure \ref{fig:setup} shows the experimental setup.
We attach the Peltier unit and the thermocouple temperature sensor on the solenoid, and controlled the solenoid's coil temperature.
We also attached two micrometers to measure and fix the solenoid stroke, the position of the solenoid's plunger.
We used the microcontroller of Arduino UNO to generate the driving PWM waveform and the current measurement.
We measured $v_0$ and $v_1$ for PWM duty cycle, $r$ of 10[\%] to 90[\%] with step of 10[\%], the temperature, $T$, of 25[\degC] to 45[\degC] with step of 5[\degC], for each solenoid.

\begin{table}[t]
\caption{Results of Training CNN}
\label{tab:train-result}
\begin{center}
\begin{tabular}{|c|cc|}\hline
             & \multicolumn{2}{|c|}{MSE} \\
Solenoid     & Position[mm] & Temperature[\degC] \\ \hline
CBS07300380  & 0.0064 & 0.1603 \\
CB10370140   & 0.1211 & 0.2013 \\
SSBH-0830-01 & 0.1528 & 0.3136 \\ \hline
\end{tabular}
\end{center}
\end{table}

We trained the CNN in Fig.\ref{fig:cnn-structure} to predict $(x, T)$ for given $(v_0, v_1, r)$ with epoch of 1,000 for each solenoid using the measured data.

Table \ref{tab:train-result} shows the training results, where we expect to predict $x$ and $T$ with errors of about 0.3[mm] and 0.5[\degC], respectively.

\section{Evaluation of Stroke Measurement}

We've carried out the evaluation of the stroke measurement using the trained CNN.
Three solenoids in Tab.\ref{tab:solenoids} are used for evaluation.
Solenoids are fixed by the fixture in Fig.\ref{fig:setup}.
Each solenoid is fixed with the plunger's position from 0[mm] to the maximum stroke with step of 1[mm].
$v_0$ and $v_1$ are measurement by ArduinoUNO, and passed to another microcontroller, M5Stack Core2\cite{Core2}, with MPU of ESP32 running at 240MHz, to perform the prediction of the position and the temperature prediction using trained CNN.
The measurement is carried out for 100 samples, for each PWM On-time, \Ton, from 1[ms] to 9[ms] with step of 1[ms].
The time of the CNN's prediction operation was measured as approximately 1[ms].
The temperature of the solenoid is measured by K type thermocouple temperature sensor.

Table \ref{tab:evaluation} shows the evaluation results, where $e_x$ and $e_T$ are the mean error of the predicted stroke position and temperature, respectively, against the fixed or measured values, for each \Ton\ and the temperature, for each solenoid.
Note that the data of SSBH-0830-01 at $T$=40[\degC] are missing due to experimental restrictions, will be measured in our future works.

The averaged prediction error of the stroke position for all \Ton\ and the temperature are around 0.2mm[mm].
The averaged prediction error of the temperature of CBS07300380, around 5[\degC] is larger than those of the other two solenoids, around 1.5[\degC].

\begin{table*}
\caption{Evaluation results of the prediction error of the stroke position ($e_x$) and the temperature ($e_T$) using the trained CNN. (a)CBS07300380, (b)CB10370140, and (c)SSBH-0830-01.}
\label{tab:evaluation}
\begin{center}
(a)\\
\begin{tabular}{|c||ccccc|c||ccccc|c|}\hline
         & \multicolumn{6}{|c||}{$e_x$[mm]} & \multicolumn{6}{|c|}{$e_T$[\degC]} \\
\Ton[ms] & $T=25$[\degC] & $T=30$[\degC] & $T=35$[\degC] & $T=40$[\degC] & $T=45$[\degC] & Avg & $T=25[$\degC] & $T=30$[\degC] & $T=35$[\degC] & $T=40$[\degC] & $T=45$[\degC] & Avg \\ \hline
1 & -0.164  & -0.073  & -0.204  & -0.170  & -0.188  & -0.160  & -3.883  & -2.616  & -3.147  & -2.498  & -2.301  & -2.889 \\
2 & -0.107  & -0.017  & -0.143  & -0.145  & -0.154  & -0.113  & -2.940  & -2.140  & -2.810  & -3.148  & -3.924  & -2.993 \\
3 & -0.089  & -0.027  & -0.157  & -0.149  & -0.172  & -0.119  & -4.678  & -3.233  & -4.252  & -4.259  & -4.392  & -4.163 \\
4 & -0.061  & 0.003  & -0.120  & -0.128  & -0.159  & -0.093  & -3.804  & -2.632  & -4.169  & -4.038  & -4.534  & -3.835 \\
5 & -0.050  & -0.003  & -0.129  & -0.140  & -0.169  & -0.098  & -3.827  & -2.897  & -4.471  & -5.025  & -5.873  & -4.419 \\
6 & -0.086  & 0.000  & -0.110  & -0.110  & -0.135  & -0.088  & -5.071  & -3.968  & -5.480  & -6.315  & -7.456  & -5.658 \\
7 & -0.092  & 0.006  & -0.088  & -0.082  & -0.379  & -0.127  & -6.263  & -4.778  & -5.383  & -6.395  & 1.170  & -4.330 \\
8 & -0.152  & -0.007  & -0.078  & -0.079  & -0.081  & -0.079  & -7.152  & -5.649  & -6.617  & -6.260  & -6.937  & -6.523 \\
9 & -0.090  & 0.051  & -0.060  & -0.055  & -0.086  & -0.048  & -8.399  & -5.989  & -6.437  & -6.277  & -7.790  & -6.979 \\ \hline
Avg & -0.099  & -0.007  & -0.121  & -0.118  & -0.169  & -0.103  & -5.113  & -3.767  & -4.752  & -4.913  & -4.671  & -4.643 \\ \hline
\end{tabular}
\vspace*{3mm}\\
(b)\\
\begin{tabular}{|c||ccccc|c||ccccc|c|}\hline
         & \multicolumn{6}{|c||}{$e_x$[mm]} & \multicolumn{6}{|c|}{$e_T$[\degC]} \\
\Ton[ms] & $T=25$[\degC] & $T=30$[\degC] & $T=35$[\degC] & $T=40$[\degC] & $T=45$[\degC] & Avg & $T=25[$\degC] & $T=30$[\degC] & $T=35$[\degC] & $T=40$[\degC] & $T=45$[\degC] & Avg \\ \hline
1 & 0.116  & 0.096  & -0.067  & 1.481  & -0.028  & 0.319  & 10.185  & -1.463  & -2.611  & 11.291  & -7.659  & 1.949 \\
2 & 0.610  & -0.062  & 0.535  & -0.145  & -0.093  & 0.169  & 10.676  & -1.364  & 8.011  & -2.064  & -3.513  & 2.349 \\
3 & -0.225  & 0.058  & -0.083  & 0.040  & 0.128  & -0.016  & -0.790  & -0.446  & -0.626  & 0.960  & 0.556  & -0.069 \\
4 & -0.197  & -0.016  & -0.197  & -0.057  & 0.032  & -0.087  & -1.392  & -0.499  & -0.570  & 1.964  & 2.291  & 0.359 \\
5 & 0.043  & 0.012  & -0.182  & -0.110  & -0.071  & -0.061  & -0.978  & -0.525  & -0.378  & 2.629  & 3.322  & 0.814 \\
6 & 0.192  & 0.143  & -0.087  & -0.144  & -0.145  & -0.008  & -0.542  & 0.094  & 0.182  & 3.106  & 4.249  & 1.418 \\
7 & 0.261  & 0.180  & -0.038  & -0.138  & -0.226  & 0.008  & -0.740  & 0.042  & 0.207  & 3.038  & 4.209  & 1.351 \\
8 & 0.401  & 0.146  & 0.010  & -0.281  & -0.486  & -0.042  & -0.499  & -0.149  & 19.564  & 3.061  & 4.194  & 5.234 \\
9 & 1.200  & 0.476  & 0.115  & -0.314  & -0.588  & 0.178  & -0.069  & -0.007  & -0.320  & 2.418  & 3.178  & 1.040 \\ \hline
Avg & 0.267  & 0.115  & 0.001  & 0.037  & -0.164  & 0.051  & 1.761  & -0.480  & 2.607  & 2.934  & 1.203  & 1.605 \\ \hline
\end{tabular}
\vspace*{3mm}\\
(c)\\
\begin{tabular}{|c||ccccc|c||ccccc|c|}\hline
         & \multicolumn{6}{|c||}{$e_x$[mm]} & \multicolumn{6}{|c|}{$e_T$[\degC]} \\
\Ton[ms] & $T=25$[\degC] & $T=30$[\degC] & $T=35$[\degC] & $T=40$[\degC] & $T=45$[\degC] & Avg & $T=25[$\degC] & $T=30$[\degC] & $T=35$[\degC] & $T=40$[\degC] & $T=45$[\degC] & Avg \\ \hline
1 & 0.031  & 0.101  & 0.156  &  & 0.136  & 0.106  & 7.045  & 2.686  & -1.276  &  & -7.406  & 0.262 \\
2 & 0.044  & 0.109  & 0.180  &  & 0.206  & 0.135  & 6.572  & 2.340  & -1.492  &  & -7.329  & 0.023 \\
3 & 0.017  & 0.061  & 0.145  &  & 0.157  & 0.095  & 3.642  & 1.404  & -1.074  &  & -5.263  & -0.323 \\
4 & -0.065  & 0.097  & 0.207  &  & 0.338  & 0.144  & -0.558  & -0.241  & -1.169  &  & -1.834  & -0.951 \\ 
5 & 0.062  & 0.091  & 0.133  &  & 0.151  & 0.109  & -1.491  & -1.301  & -1.779  &  & -2.324  & -1.724 \\
6 & 0.193  & 0.153  & 0.163  &  & 0.089  & 0.149  & -1.752  & -1.731  & -2.329  &  & -3.034  & -2.212 \\
7 & 0.448  & 0.345  & 0.306  &  & 0.170  & 0.317  & -2.094  & -1.999  & -2.384  &  & -2.860  & -2.334 \\
8 & 0.410  & 0.401  & 0.364  &  & 0.307  & 0.371  & -2.532  & -2.556  & -3.080  &  & -3.221  & -2.847 \\
9 & 0.589  & 0.898  & 0.414  &  & 0.030  & 0.482  & -1.783  & -4.772  & -3.375  &  & -3.743  & -3.418 \\ \hline
Avg & 0.192  & 0.251  & 0.230  &  & 0.176  & 0.212  & 0.783  & -0.685  & -1.995  &  & -4.112  & -1.503 \\ \hline
\end{tabular}
\end{center}

\end{table*}

\section{Preliminary Experiment of Stroke Control}

We also carried out the preliminary experiments of stroke control using the prediction CNN.
The mechanical parameters are not considered in this control algorithm, where the simple PID control algorithm using the target stroke position and the predicted position for evaluating position prediction.

We used M5Stack Core2 to control the solenoid, as well as the position prediction using CNN from the measured $v_0$ and $v_1$.
The predicted position, $x_P$ is calculated in main loop using the CNN from the measured $v_0$ and $v_1$, and the driving PWM ON time, $T_1$.
The PWM ON time at next step, $T_1'$ is calculated as follows based on simple PID control, with the target position of $x_T$.
\begin{equation}
  T_1' = T_1 + K_p (x_P - x_T)
\end{equation}
The P gain, $K_p$, is determined as 10.0 from trials in this experiment.

The actual stroke position is measured using the laser displacement meter, Keyence LK-H050, whose accuracy is 0.025[$\mu$m] with sampling rate of 1[kHz].

\begin{figure}
\begin{center}
\includegraphics[width=0.9\columnwidth]{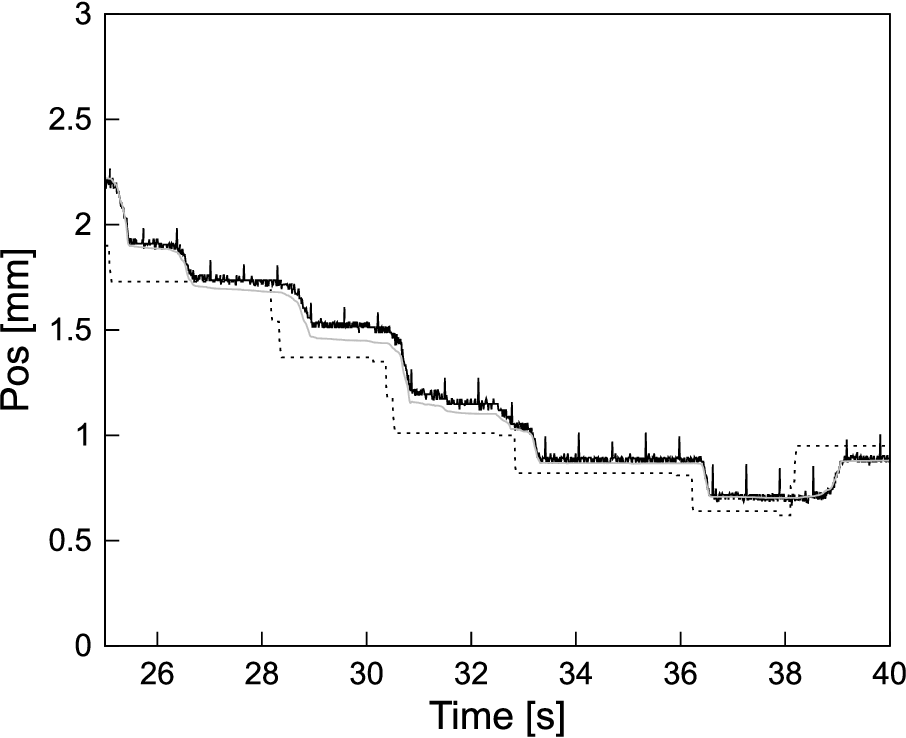} \\
(a) \\
\vspace*{5mm}
\includegraphics[width=0.9\columnwidth]{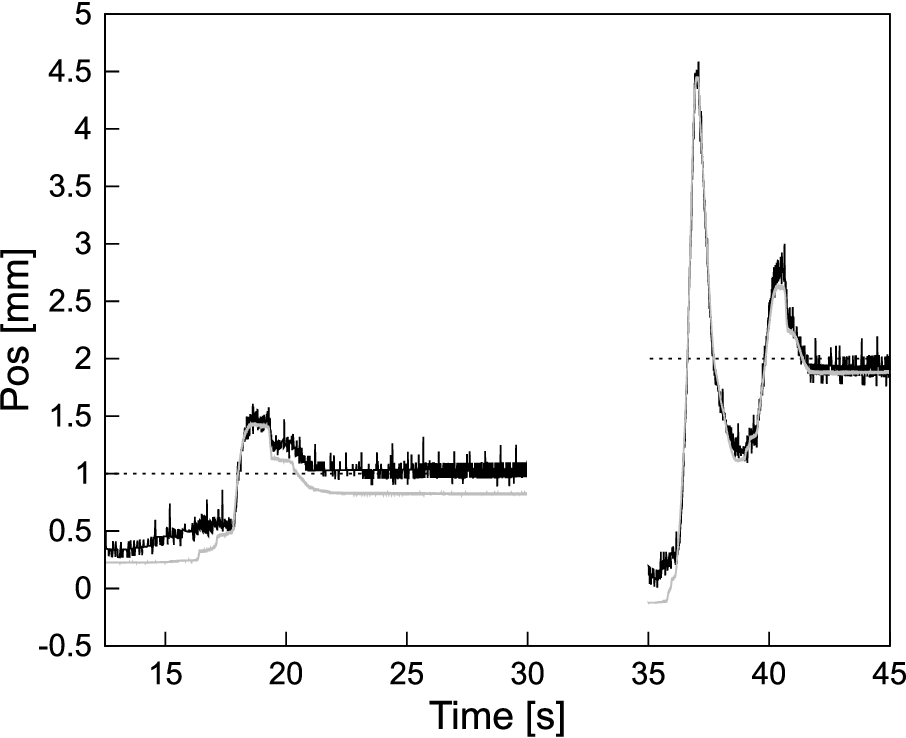}\\
(b) \\
\vspace*{5mm}
\includegraphics[width=0.9\columnwidth]{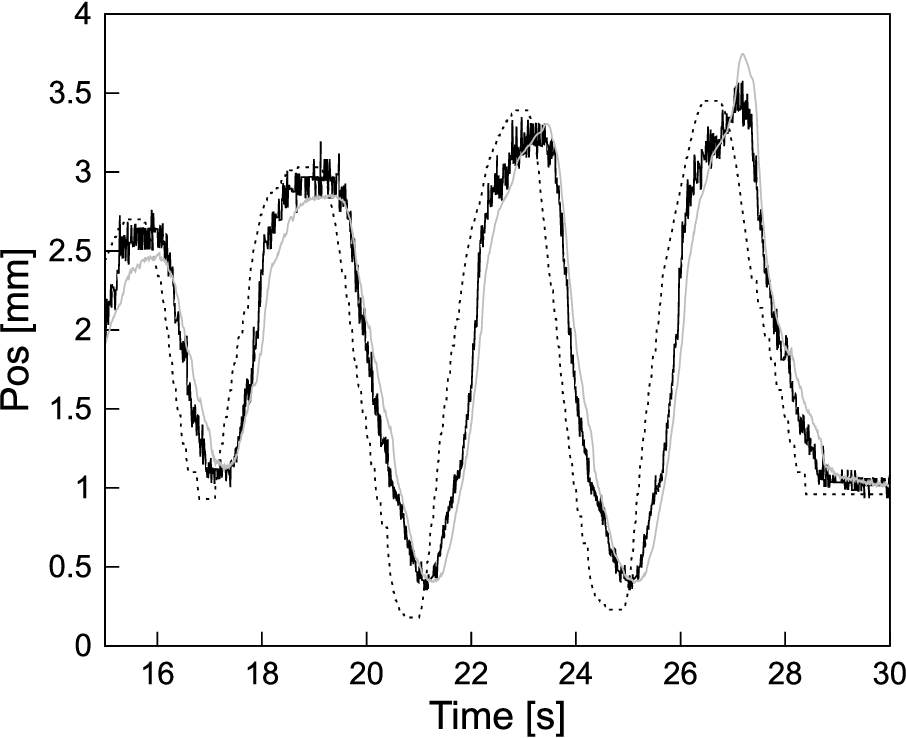}
(c) \\
\end{center}
\caption{Experimental results of stroke control. Dotted line: Target position, Solid line: predicted position, Gray line: measured position. (a)CBS07300380, (b)CB10370140, and (c)SSBH-0830-01.}
\label{fig:experiment-control}
\end{figure}

\begin{figure}
\begin{center}
\includegraphics[width=0.9\columnwidth]{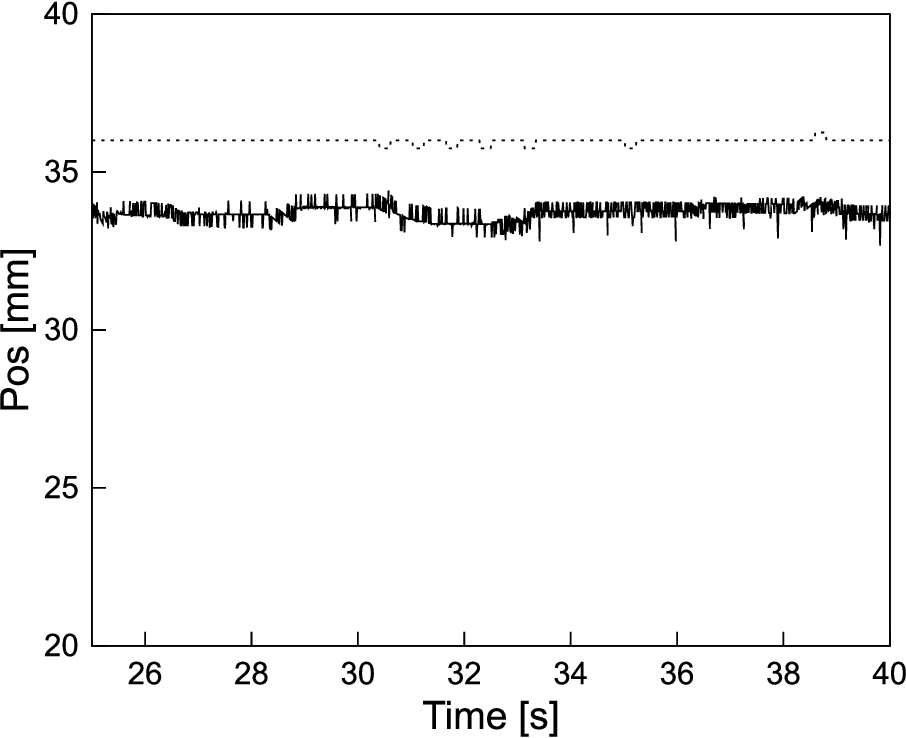} \\
(a) \\
\vspace*{5mm}
\includegraphics[width=0.9\columnwidth]{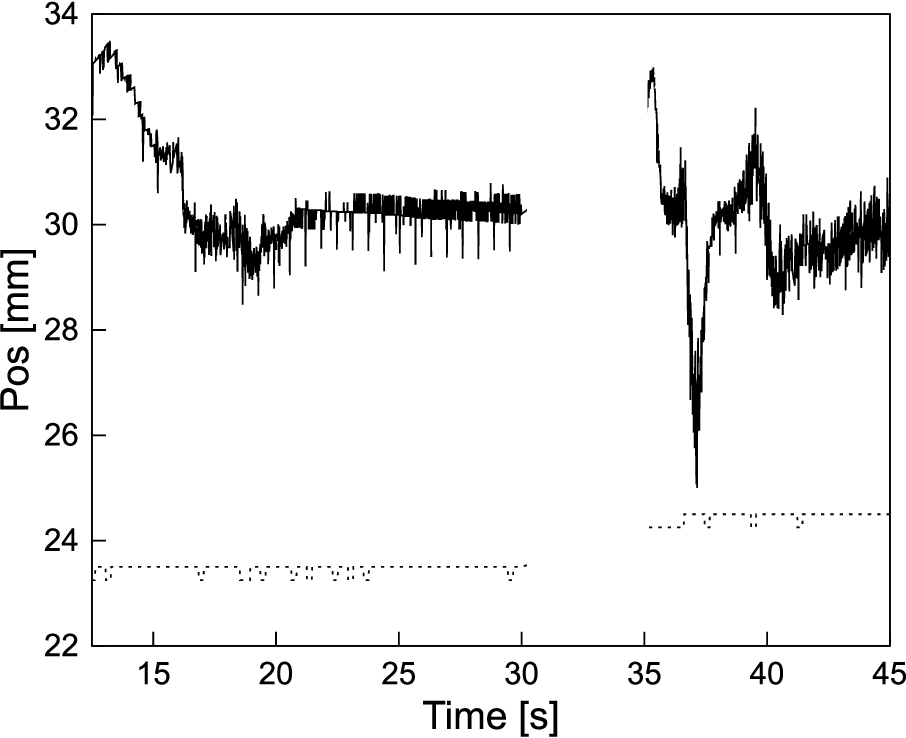}\\
(b) \\
\vspace*{5mm}
\includegraphics[width=0.9\columnwidth]{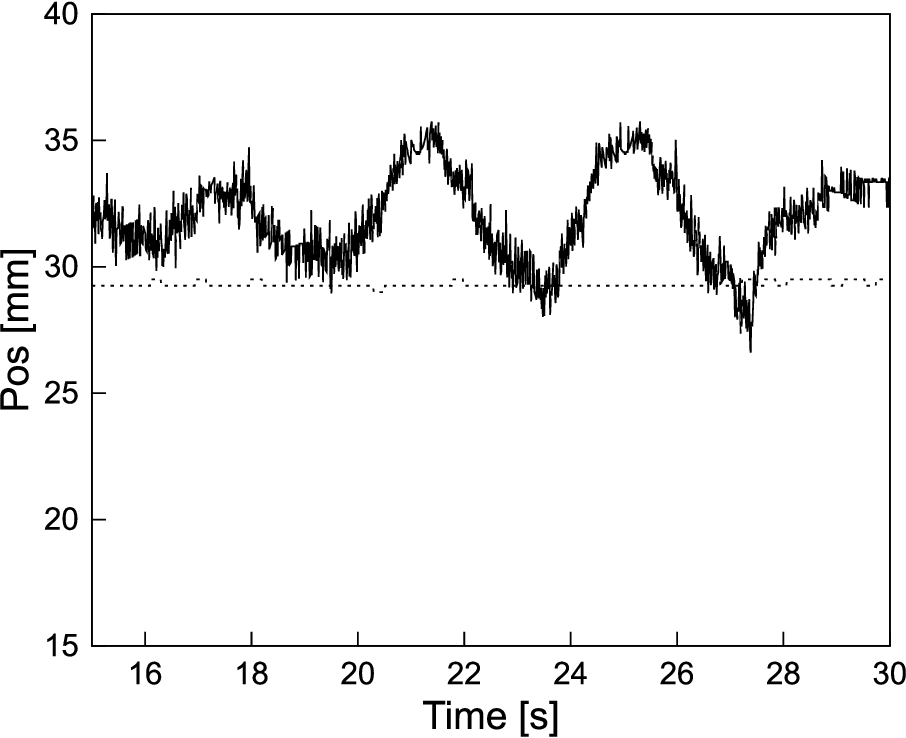}
(c) \\
\end{center}
\caption{Experimental results of temperature prediction in control of Fig.\ref{fig:experiment-control}.Dotted line: Measured temperature, Solid line: predicted temperature. (a)CBS07300380, (b)CB10370140, and (c)SSBH-0830-01.}
\label{fig:experiment-temperature}
\end{figure}

Figure \ref{fig:experiment-control} shows the experimental results of the stroke position control.
Note that in Fig.\ref{fig:experiment-control}(b) has lack of data between 30[s] and 35[s] due to the restriction of experiments of setting the target position.

As shown in Fi.\ref{fig:experiment-control}, the position tracking for the target position can be achieved with less error than 0.2[mm]
However, we also find some overshoots at the target position changes, and can also find an oscillation where the target position is close to the ``pulled'' state of the plunger.
Solenoid itself has the mechanical characteristics that the plunger is rapidly ``pulled'' at close position of ``pulled'' state, and it affects the PID control gains, and makes the position control more sensitive.

Figure \ref{fig:experiment-temperature} shows the predicted and the measured temperature of the solenoid in control of Fig.\ref{fig:experiment-control}.
The CNN model's training result indicates that the prediction error of the temperature is larger than that of the position.
The experimental result in Fig.\ref{fig:experiment-temperature} shows the prediction error of the temperature in Fig.\ref{fig:experiment-temperature}(a) indicates the error around 2[\degC], while it depends on the position in Fig.\ref{fig:experiment-temperature}(c).

\section{Conclusion and Future Works}

In this paper, we proposed to method to measure the stroke and the temperature of the solenoid using convolution neural network (CNN) for two measured values of driving current, as well as its evaluation on the accuracy.
The measured averaged prediction error of the stroke position and the temperature are around 0.2mm[mm] and 1.5[\degC], respectively.

We also demonstrated the preliminary experiment of the stroke control using the proposed method with considering only the stroke position.
Although the position tracking with less error than 0.2[mm] is achieved, there are situation of the overshoots and the oscillations during position control.
The detailed control model with physical parameters and the generated force of the solenoid and its load will be discussed in our future works for improving the control stability and accuracy.
The detailed discussion on the temperature prediction error are also our future works.


\begin{biography}
\profile{n}{Junichi Akita}{%
He received B.S., M.S. and Ph.D. degrees in electronics engineering from the University of Tokyo, Japan in 1993, 1995 and 1998 respectively. He joined the Department of Computer and Electrical Engineering, Kanazawa University as a research associate in 1998. He moved to the Department of Media Architecture, Future University Hakodate as an assistant professor in 2000. He moved to the Department of Information and Systems Engineering, Kanazawa University as an assistant professor in 2004. From 2022, he is a professor at School of Transdisplenary Science for Innovation, Kanazawa University. He is also interested in electronics systems including VLSI systems in the applications of human-machine interaction and human interface. He is a member of the Institute of Electronics, Information and Communication Engineers of Japan, Information Processing Society of Japan, and the Institute of Image Information and Television Engineering.}
\end{biography}

\end{document}